\documentclass[aps,prl,reprint, superscriptaddress]{revtex4-1}

\usepackage[pdftex]{graphicx}

\begin{document}

\title{H.E.S.S. limits on line-like dark matter signatures in the 100 GeV to 2 TeV energy range close to the Galactic Centre}

\author{H.E.S.S. Collaboration}
\noaffiliation

\author{H.~Abdalla}
\affiliation{Centre for Space Research, North-West University, Potchefstroom 2520, South Africa}

\author{A.~Abramowski}
\affiliation{Universit\"at Hamburg, Institut f\"ur Experimentalphysik, Luruper Chaussee 149, D 22761 Hamburg, Germany}

\author{F.~Aharonian}
\affiliation{Max-Planck-Institut f\"ur Kernphysik, P.O. Box 103980, D 69029 Heidelberg, Germany}
\affiliation{Dublin Institute for Advanced Studies, 31 Fitzwilliam Place, Dublin 2, Ireland}
\affiliation{National Academy of Sciences of the Republic of Armenia,  Marshall Baghramian Avenue, 24, 0019 Yerevan, Republic of Armenia}

\author{F.~Ait Benkhali}
\affiliation{Max-Planck-Institut f\"ur Kernphysik, P.O. Box 103980, D 69029 Heidelberg, Germany}

\author{A.G.~Akhperjanian}
\affiliation{National Academy of Sciences of the Republic of Armenia,  Marshall Baghramian Avenue, 24, 0019 Yerevan, Republic of Armenia}
\affiliation{Yerevan Physics Institute, 2 Alikhanian Brothers St., 375036 Yerevan, Armenia}

\author{T.~Andersson}
\affiliation{Department of Physics and Electrical Engineering, Linnaeus University,  351 95 V\"axj\"o, Sweden}

\author{E.O.~Ang\"uner} 
\affiliation{Institut f\"ur Physik, Humboldt-Universit\"at zu Berlin, Newtonstr. 15, D 12489 Berlin, Germany}

\author{M.~Arrieta}
\affiliation{LUTH, Observatoire de Paris, PSL Research University, CNRS, Universit\'e Paris Diderot, 5 Place Jules Janssen, 92190 Meudon, France}

\author{P.~Aubert}
\affiliation{Laboratoire d'Annecy-le-Vieux de Physique des Particules, Universit\'{e} Savoie Mont-Blanc, CNRS/IN2P3, F-74941 Annecy-le-Vieux, France}

% ...

\date{March 30, 2016}%

\begin{abstract}
A search for dark matter line-like signals was performed in the vicinity of the Galactic Centre by the H.E.S.S. experiment on observational data taken in 2014. An unbinned likelihood analysis was developed to improve the sensitivity to line-like signals. The upgraded analysis along with newer data extend the energy coverage of the previous measurement down to 100~GeV. The 18~h of data collected with the H.E.S.S. array allow one to rule out at 95\%~CL the presence of a 130~GeV line (at $l = -1.5^{\circ}, b = 0^{\circ}$ and for a dark matter profile centred at this location) previously reported in {\it{Fermi}}-LAT data. This new analysis overlaps significantly in energy with previous {\it{Fermi}}-LAT and H.E.S.S. results. No significant excess associated with dark matter annihilations was found in the energy range 100~GeV to 2~TeV and upper limits on the gamma-ray flux and the velocity weighted annihilation cross-section are derived adopting an Einasto dark matter halo profile. Expected limits for present and future large statistics H.E.S.S. observations are also given.

\end{abstract}

\maketitle

\section{Introduction}
 
Weakly interacting massive particles (WIMPs) are among the most studied candidates to explain the longstanding elusive nature of dark matter (DM) and have been the target of a large number of searches (see \cite{Bertone:2004pz} for a review). In particular, the indirect detection of DM using gamma rays is considered one of the most promising avenues as it can probe both its particle properties and distribution in the universe WIMP annihilations produce a continuum energy spectrum of gamma rays up to the DM mass as well as one or several gamma-ray lines. Although the fluxes of such mono-energetic features are mostly suppressed compared to the continuum, a line spectrum is easier to distinguish in regions of the sky with high astrophysical gamma-ray backgrounds~\cite{bib:Conrad}. 

A previous search for line signatures using H.E.S.S. in phase I (H.E.S.S.~I) has been published~\cite{hess_line} with 112~h of observation time. As no significant excess was found, the study presented upper limits on the flux and velocity-averaged annihilation cross-section $\langle\sigma v\rangle$ at the level of $10^{-6} \mathrm{m}^{-2}\mathrm{s}^{-1}\mathrm{sr}^{-1}$ and $10^{-27} \mathrm{cm}^{3}\mathrm{s}^{-1}$ for WIMP masses between 500~GeV and 20~TeV. The space-borne {\it{Fermi}} Large Area Telescope ({\it{Fermi}}-LAT)~\cite{Atwood:2009ez} was until recently the only instrument capable of probing a DM induced gamma-ray line signal in the direction of the Galactic Centre of around 100 GeV in energy. Analyses based on public data have found indications of an excess signal at around 130 GeV in the vicinity of the Galactic Centre finding a best fit position for the centroid of the excess at ($l = -1.5^{\circ}, b = 0^{\circ}$) \cite{Bringmann:2012vr,bib:4,su_fink,Bringmann:2012ez}. Later, revised analyses of the {\it{Fermi}}-LAT team found background-compatible results~\cite{Ackermann:2013uma,Ackermann:2015lka}. In order to resolve the controversy with an independent measurement, the H.E.S.S. collaboration performed dedicated observations of the Galactic Centre vicinity using its newly commissioned fifth telescope. The larger effective area and lower energy threshold allow to eliminate the energy gap between previously reported {\i{Fermi}}-LAT and H.E.S.S.~I results.

The present paper is organised as follows: first the H.E.S.S. experiment and event reconstruction are briefly described, then the analysis method is discussed, followed by the presentation of the results and concluding remarks.

\section{H.E.S.S. experiment and Line Scan event reconstruction}

The H.E.S.S. experiment \cite{Aharonian:2006pe} covers a wide range of astrophysical and fundamental physics topics, including indirect DM searches. Between 2002 and 2012, H.E.S.S. consisted of four 12~m diameter telescopes (CT1-4). A fifth telescope (CT5) with a larger mirror diameter of 28~m and newly designed camera~\cite{bolmont} augmented the array in 2012, reducing the energy threshold significantly to below 100~GeV. This array configuration constitutes H.E.S.S. phase 2 (H.E.S.S.~II). H.E.S.S. triggers on two different types of events: monoscopic single-telescope events from CT5 and stereoscopic CT1-5 events. The former exclusively rely on the information from CT5, whereas the latter require at least two telescopes to record an individual shower. In the standard observation mode, both monoscopic and stereoscopic events are recorded at the same time and CT5 participates in more than 95\% of the events that are triggered by more than one telescope.

Throughout the past years, several existing H.E.S.S. analysis chains have been extended to reconstruct monoscopic events and those recorded with two different types of telescopes~\cite{HollerICRC,MurachICRC,refImpactMono,HollerCrabICRC,ParsonsSgrAICRC}. The search for a gamma-ray line feature around 130 GeV requires a selection of event cuts that allows for a reasonably low energy threshold and an excellent energy resolution. For this purpose the reconstruction technique described in~\cite{refModel,HollerICRC} has been chosen with stereoscopic events considered in the analysis. An analysis with monoscopic events (CT5 only)~\cite{refImpactMono} has also been prepared as a cross-check, which we describe later. To efficiently suppress the charged cosmic-ray background, analysis requirements have been defined and tested on a-priori independent data sets obtained from observations of standard calibration sources such as PKS 2155-304 or the Crab nebula. The chosen configuration of event cuts for this analysis setup achieves the desired low energy threshold of 80~GeV, a better background rejection efficiency than for monoscopic events and an excellent relative energy resolution of 14\% for gamma rays of energies below 300~GeV.

Due to uncertainty in the position of the 130 GeV excess, the H.E.S.S.~II observations were implemented in a scanning mode of the Galactic plane, with pointing positions ranging from -$2.5^{\circ}$ to $0.5^{\circ}$ in longitude $l$ in steps of $0.7^{\circ}$ and at $b = \pm 0.8^{\circ}$. A total of 18~h of data have been accumulated from April to July 2014: 2.8 h were used to choose the event reconstruction mode, for the studies related to the background Probability Density Function (\textit{PDF}) determination, employed in the likelihood fit, and the study of systematic effects. The remaining 15.2~h were used for the final results for the gamma-ray line DM signal search between 100~GeV and 2~TeV. Data quality checks were performed based on the global array and the individual telescope status. Cuts have been applied on the telescope trigger rates, trigger rate stability and the broken pixel fraction of the camera. The resulting data sample covers observations at zenith angles ranging from $10^{\circ}$ to $30^{\circ}$. Gamma-ray candidate events passing all of the aforementioned cuts and falling into either the signal region (ON-source) or in any of the defined background control regions (OFF-source) are then utilised in the likelihood-based line-search analysis as described in the next section.

\section{Analysis methodology}
The results presented in this paper were obtained with a likelihood fit of the line-like signal in the ON-source region with modelling of the background contribution with OFF-source data. The fit was performed using an event-by-event likelihood procedure optimised for DM searches in the Galactic Centre region. Here no background subtraction was performed in order to preserve maximal sensitivity to the DM signal. Since measured energy distributions were considered in the likelihood fit, there is no need for acceptance corrections on the background measured spectra, strongly limiting the associated systematic uncertainties which are discussed in the section presenting the results.  Additional systematic uncertainties may be introduced by night sky background differences between the background control and signal regions. To minimise these uncertainties, the OFF-source regions associated directly with a given ON-source position were chosen close to the ON-source region. The measured energy distributions in these OFF-source regions were used for the construction of the background \textit{PDF}, a major component in the likelihood discussed below.

The likelihood function is composed of a Poisson normalisation term (based on the total number of events in the signal and background regions) and a spectral term related to the expected spectral contribution of the signal and the background component in the analysis region of interest (ROI). A description of this approach, called full likelihood method below, is given in \cite{bib:Aleksic}. 

The likelihood formula reads as:

\small
\begin{eqnarray}
&\mathcal{L}(N_{signal},N_{bckg}|N_{ON},N_{OFF},E_{i}) = \nonumber \\ &\frac{(N_{signal}+N_{bckg})^{N_{ON}}}{N_{ON}!}e^{-(N_{signal}+N_{bckg})} \times \frac{(\alpha N_{bckg})^{N_{OFF}}}{N_{OFF}!}e^{-\alpha N_{bckg}} \nonumber \\ &\times \prod \limits_{i=1}^{N_{ON}} \Big( \eta \times PDF_{signal} (E_{i}) + (1-\eta) \times PDF_{bckg} (E_{i}) \Big)
\label{eq:fulllikelihoodformula}
\end{eqnarray}
\normalsize
where $N_{ON}$ and $N_{OFF}$ are the measured number of events in the signal and background regions, $\alpha$ the exposure ratio between background and signal regions, $E_{i}$ (with $i \in [1,N_{ON}]$) representing a vector of energies of events measured in the signal region, and $\eta = {N_{signal}}/({N_{signal}+N_{bckg}})$ is the line signal fraction in the ON region sample. $\textit{PDF}_{signal}$ and $\textit{PDF}_{bckg}$ are the probability density functions for the signal and background components that refer to measured energy spectra, that is, photon energies smeared by the Instrument Response Functions (IRFs). The $\textit{PDF}_{signal}$ is obtained from dedicated mono-energetic gamma-ray simulations of signals for each DM particle mass considered in the analysis. The $\textit{PDF}_{bckg}$ corresponds to the best fit of the normalised energy distribution of events reconstructed in the OFF regions. No additional term corresponding to the fit of the $\textit{PDF}_{bckg}$ was added to the likelihood formula~(\ref{eq:fulllikelihoodformula}). The number of signal ($N_{signal}$) and background ($N_{bckg}$) events are free parameters of the model, while additional information on the signal and background spectral shape is included in the fit. The line energy position $E_{line}$ is kept fixed, and the line signal fraction $\eta$ which represents the relative contribution of the signal in the analysed region is fitted.

The IRFs were obtained from the full gamma-ray MC simulations of the gamma-ray showers and of the H.E.S.S. instrument. They were employed in the dedicated MC simulations to derive the expected measured energy distributions leading to $\textit{PDF}_{signal}$ and $\textit{PDF}_{bckg}$. An optimal circular signal region of $0.4^{\circ}$ radius was found using the method of Rolke et al.,~\cite{bib:Rolke}, corresponding to a solid angle of $\Delta \Omega = 1.531 \times 10^{-4}$ sr.

The resulting sensitivity estimates computed with MC simulations for a line scan between 100~GeV and 2~TeV as well as the 95\% confidence level (CL) limits derived from the data sample are presented below. 

\section{Results}

\begin{figure}
 \includegraphics[width=0.4\textwidth]{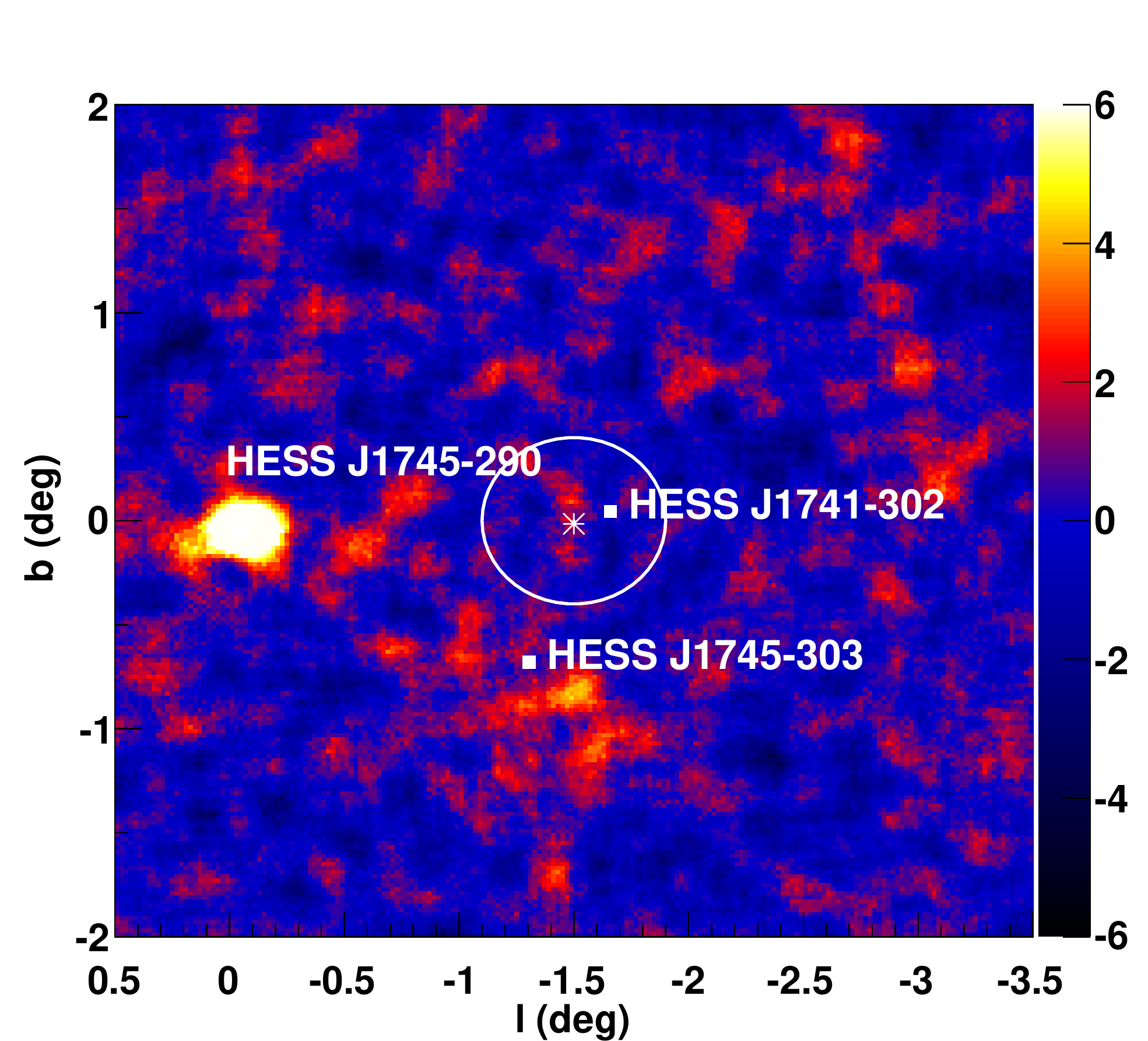}
 \includegraphics[width=0.42\textwidth]{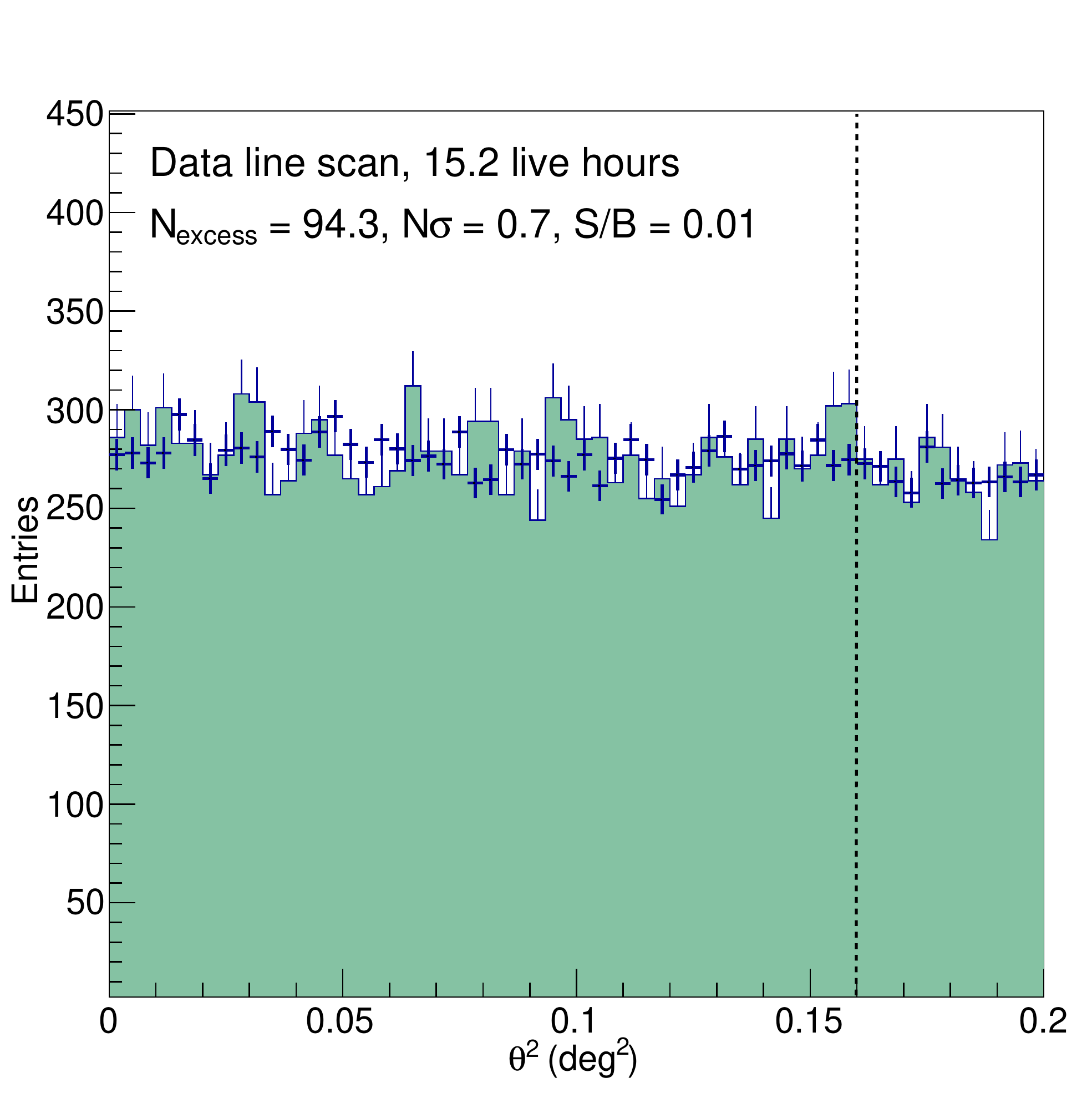} 
 \caption{Significance map presented in Galactic coordinates ({\it{top}}) and emission angle square ($\theta^2$) distribution ({\it{bottom}}) in the considered ROI. The ROI is expressed in the map with a white circle centred on the 130 GeV excess ($-1.5^{\circ}$, $0^{\circ}$) marked with a white cross. The known source HESS J1745-290 is detected, even at large angular offset. The dashed vertical line in the bottom shows the $\theta$ cut of $0.4^{\circ}$.}
 \label{fig:linescan_map}
\end{figure}

At first, a search for an excess in the ON-source region was performed by using OFF-region empty field data. It should be noted that despite the signal region being displaced from the Galactic Centre (GC) position, the 130 GeV excess ROI may still be subject to contributions from surrounding astrophysical sources. In particular, the bright extended source HESS J1745-303~\cite{aharonian2008exploring} was excluded (a mask of $0.4^{\circ}$) while the contribution from HESS J1741-302~\cite{tibolla2009new} was estimated to be negligible. The significance map shown in Figure~\ref{fig:linescan_map} was reconstructed with an annular background region~\cite{Aharonian:2006pe} around the signal region for the 15.2~h data set. In the absence of any genuine gamma-ray signal in the field-of-view, the significances derived from background fluctuations follow a Gaussian distribution with a width of one, as it is the case once the significant excess at the position of HESS J1745-290 \cite{2010MNRAS.402.1877A} is excluded, coincident with the supermassive black hole SgrA$^*$. As also shown in Figure \ref{fig:linescan_map}, no significant excess ($N_{signal}$) was found in the $0.4^{\circ}$ radius ROI at the best-fit position of the 130 GeV excess $(l,b)=(-1.5^{\circ}, 0^{\circ})$. Therefore, upper limits were derived for a line-like signal in the energy range from 100~GeV to 2~TeV.
 
\noindent
\label{MC}

The number of measured background events in the ROI of $0.4^{\circ}$ and the $\textit{PDF}_{bckg}$ parametrisation were derived from the measured energy distributions in the data control OFF-source regions symmetrically surrounding the 130 GeV excess. The likelihood fits covered two pre-defined energy ranges from 80~GeV to 1~TeV and from 200~GeV to 3~TeV which allowed our observations to probe line signals with energy from 100 to 500~GeV and from 500~GeV to 2~TeV, respectively, ensuring a large energy lever arm in the fit in each case. For each line energy, upper limits on $\eta$ and subsequently on the number of excess events, $N$, were obtained using equation (\ref{eq:fulllikelihoodformula}). The $\eta^{95\% CL}$ upper-limit value was obtained from a one-sided cut on the log-likelihood function corresponding to its increase by 2.71. To derive the sensitivity expectations, we use the median of the 95\%~CL upper limits distributions obtained from a large number of simulations performed assuming 15.2 and 112~h of time exposure.

\begin{figure}[t]
 \centering
 \includegraphics[width=0.48\textwidth]{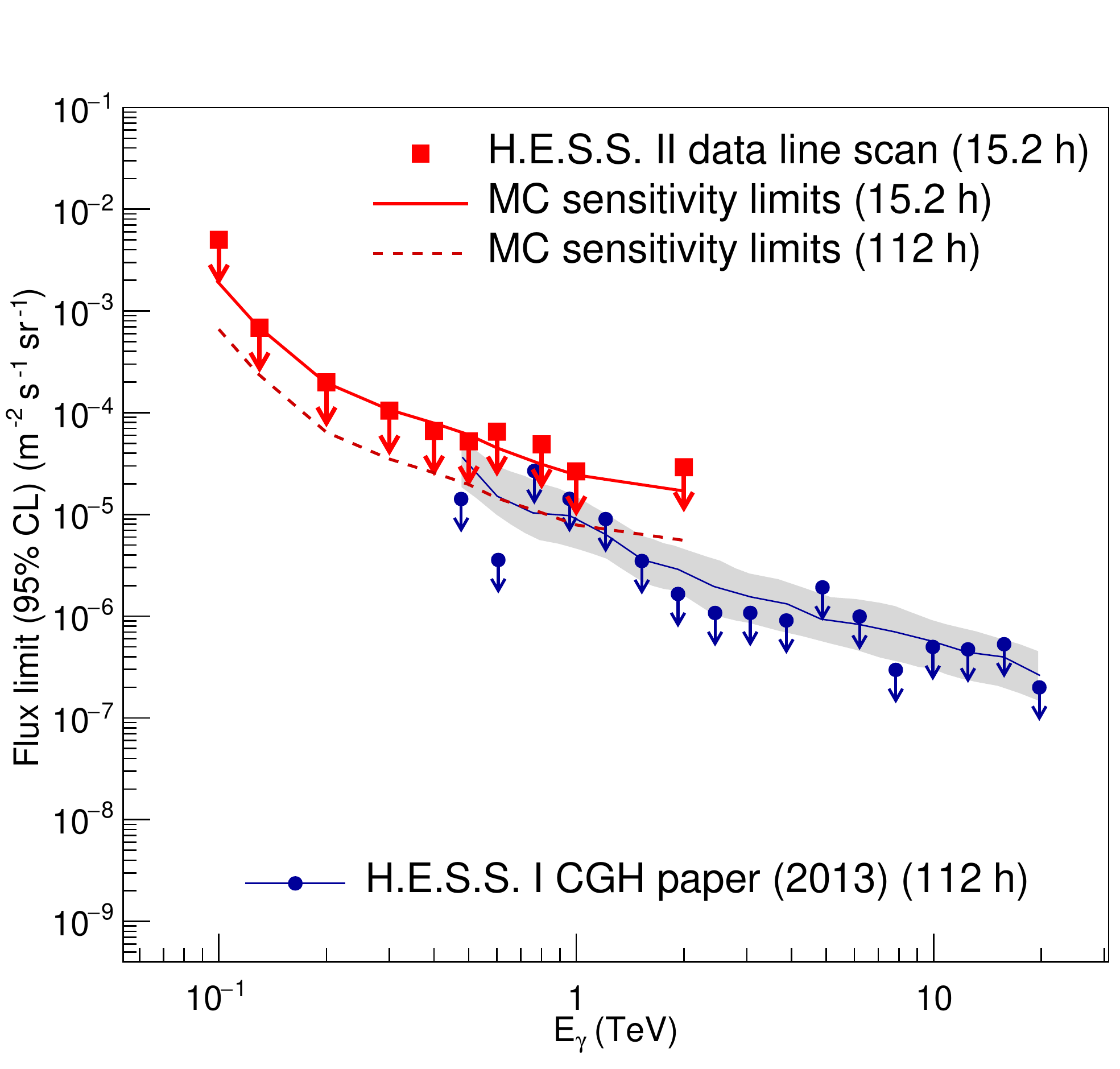}
 \caption{Flux limits at 95\%~CL for a line scan between 100~GeV and 2~TeV. The results obtained from 15.2 h of data are represented by points in red. The red dashed line represents the limits expected for 112~h of observation time, calculated as the median limits from 500 simulated data sets. The red solid line is given for 15.2~h MC simulations. Former limits from H.E.S.S.~I \cite{hess_line} obtained in the Central Galactic Halo (CGH) region are represented as blue data points (the grey band displaying the level of systematic uncertainties).}
 \label{fig:mc_limits_flux}
\end{figure}
\begin{figure}[t]
 \centering
 \includegraphics[width=0.48\textwidth]{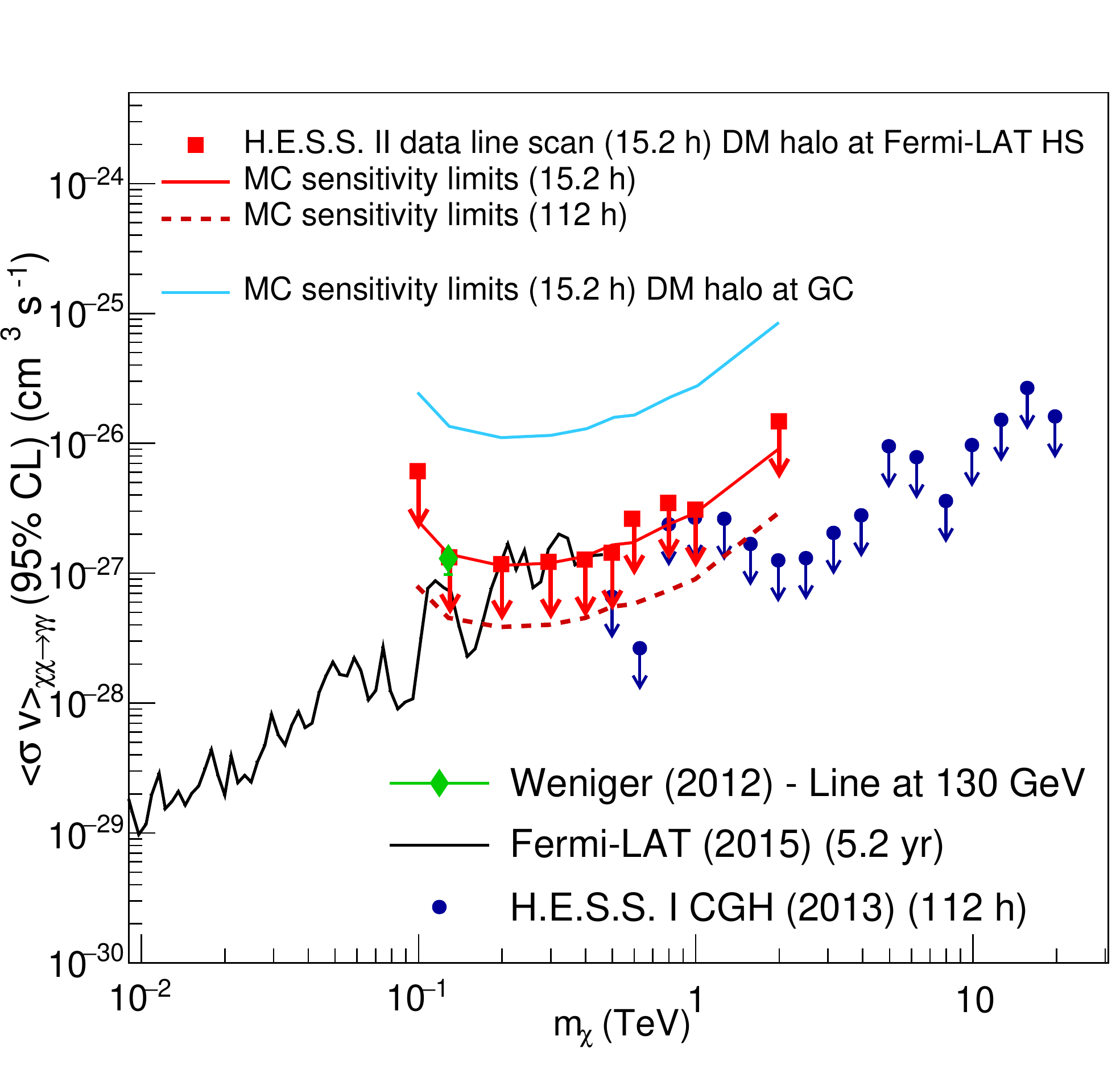}
\caption{$\langle\sigma v\rangle$ limits at 95\% CL (red points) for the line scan between 100~GeV and 2~TeV, derived from 15.2~h of data and using an Einasto DM profile with $\Phi_{astro}$ value calculated with CLUMPY package~\cite{clumpy} ($\rho_{s}$ = 20~kpc, $r_{s}$ = 0.17). 
The MC estimations are presented with the same conventions as in Figure \ref{fig:mc_limits_flux}. Former limits from H.E.S.S.~I \cite{hess_line} obtained in the CGH region and {\it{Fermi}}-LAT \cite{Ackermann:2015lka} are represented by blue and black data points, respectively. The $\langle\sigma v\rangle$ value corresponding to the 130~GeV line feature reported as R16 in \cite{bib:4} is shown in green. The limits extracted with assumption of the DM halo position at the GC are shown with a continuous blue line (see in the text). It should be noted that the comparison of the limits on the hotspot obtained in this work can not be directly done with the H.E.S.S.~I results as DM halo was centred on the Galactic Centre position in the sky. In case of {\it{Fermi}}-LAT, the red curve can still be compared to the {\it{Fermi}}-LAT limits as the latter would only be marginally modified (at the level of 1\%) by the displacement of the DM halo, given the very large size of the ROI ($16^{\circ}$ of radius) in use.}
 \label{fig:mc_limits_sigmav}
\end{figure}
\noindent

The limits on the flux ($\Phi$) and on the DM velocity averaged annihilation cross-section $\langle\sigma v\rangle$ were derived as: 

\begin{eqnarray}
 &\Phi^{95\% CL} = \frac{N_{\gamma}^{95\% CL}}{T_{OBS}} \times \frac{\int \limits_{E_{min}}^{E_{max}} {dN}/{dE_{\gamma}}(E_{\gamma})dE_{\gamma}}{\int \limits_{E_{min}}^{E_{max}} A_{eff}(E_{\gamma}) {dN}/{dE_{\gamma}}(E_{\gamma})dE_{\gamma}} \\ \nonumber \\ &\langle\sigma v\rangle^{95\% CL} = ({8 \pi m_{DM}^{2}}/{2\Phi_{astro}}) \times \Phi^{95\% CL}
\end{eqnarray}
where $T_{OBS}$ is the observation time, $A_{eff}$ and ${dN}/{dE}$ are, respectively, the effective area for gamma rays and the differential energy spectrum of the expected DM signal expressed as functions of the true energy, $m_{DM}$ the DM particle mass, $[E_{min},E_{max}]$ are the bounds of the energy range. The astrophysical factor $\Phi_{astro}$ is given by the integral of the squared DM density along the line-of-sight $l.o.s.$ and solid angle $\Omega$. A dark matter distribution following an Einasto profile \cite{einasto} with halo parameters given in \cite{hess_line} has been considered at the centre of the ROI resulting in the value of $\Phi_{astro} = 2.46 \times 10^{21}\ \textrm{GeV}^{2} \textrm{cm}^{-5}$. For DM annihilating into two gamma rays, the differential energy spectrum is ${dN}/{dE_{\gamma}} \sim 2\delta(E_{\gamma}-m_{\chi})$ where the factor of two results from the annihilation of DM particles into two photons.
 
Limits on the flux per steradian and on $\langle\sigma v\rangle$ obtained from MC simulations and those calculated with the 15.2~h of data are presented in Figures \ref{fig:mc_limits_flux} and \ref{fig:mc_limits_sigmav}, respectively, and show the potential of the applied method for the DM line signal detection. The measured limits are in good agreement with the expected sensitivity. The limits obtained with H.E.S.S.~II for a DM density profile centred on the 130 GeV excess position efficiently complement previous limits of H.E.S.S.~I \cite{hess_line} and cover the gap in mass between 300 and 500~GeV, even though H.E.S.S.~II results are derived for a different location in the sky. Due to differences in the analysis methods and a limited size of the current data sample a combination of the results obtained by H.E.S.S. phase I and phase II was not performed.

The case of the DM halo centred on the GC was also analysed and the results are shown in Figure \ref{fig:mc_limits_sigmav}, keeping the ROI on the 130 GeV excess position. The decrease in sensitivity by a factor of 8 to 10 can be explained by a decrease in the $\Phi_{astro}$ value by a factor of 4.3 ( $\Phi_{astro} = 5.6 \times 10^{20}\ \textrm{GeV}^{2} \textrm{cm}^{-5}$). In this case the DM signal leakage into the OFF regions was 40 $\%$, adding another factor of two in the total loss in sensitivity for the line search studies with data sample dedicated to the 130 GeV excess.

For the particular case of the 130~GeV excess, the likelihood method yielded the 95\%~CL limit on the line signal fraction $\eta$ of 0.0083 leading to ${N_{\gamma}^{95\% CL}}$ of 102.8 events. The 95\%~CL upper limits on the flux and $\langle\sigma v\rangle$ for data and MC simulations are summarised in Table \ref{tab:table1} for both Einasto~\cite{einasto} and Navarro-Frenk-White (NFW)~\cite{Navarro:1996gj} DM halo profiles. 

\begin{table}
 \begin{center}
  \begin{tabular}{lccc}
   \hline
   \hline
   & $\Phi^{95\% CL}/\Delta\Omega$ & $\langle\sigma v\rangle^{95\% CL}$ & $\langle\sigma v\rangle^{95\% CL}$ \\
   & $10^{-4}\ \gamma\ \textrm{m}^{-2} \textrm{s}^{-1} \textrm{sr}^{-1}$ & $10^{-27}\ \textrm{cm}^{3} \textrm{s}^{-1}$ & $10^{-27}\ \textrm{cm}^{3} \textrm{s}^{-1}$ \\
   \hline
   & & Einasto profile  & NFW profile  \\
   \hline
   Data & $8.4 $ & $ 1.38 $ & $1.43$ \\
   MC & $8.6 $ & $ 1.42 $ & $1.56$ \\
   \hline
   \hline
  \end{tabular}
 \end{center}
 \caption{95\%~CL limits on the flux (per solid angle unit) and $\langle\sigma v\rangle$ for the detection of the 130~GeV line. The limits on $\langle\sigma v\rangle$ are given for Einasto and NFW DM halo profiles. The MC values are coming from the simulations of 15.2~h of observation time. The quoted values do not include the systematic effects.}
 \label{tab:table1}
\end{table}
The cross-check studies with independent calibration and reconstruction, here in monoscopic mode, confirmed the conclusion of no significant excess at 130~GeV and the exclusion at $95\%$ CL for the 130 GeV excess. Due to the large extension of the galactic DM halo, a fraction of the expected DM signal leaks into the background regions, found to be at the level of 25\% of the DM signal in the ROI. The presented $\langle\sigma v\rangle$ limits account for this effect. The impact of various systematic uncertainties was evaluated with full MC simulations including those of radial acceptance effects within the signal region and were found to only affect the limits obtained at the few percent level. As the signal region is sufficiently large there is no effect due to the point spread function. Finally, to estimate the impact of systematic uncertainties in the limits calculation for the considered sources of errors such as IRF values, the global energy scale, the background \textit{PDF} shape and the diffuse emission component included in the background regions, nuisance parameters modelled with Gaussian functions were introduced in the full likelihood calculations. The impact of each systematic effect was studied with 500 MC simulations providing statistically calibrated results. The background \textit{PDF} shape has been identified as dominant source of systematic uncertainties, changing $95\%$~CL limits by 10 to 15 $\%$ depending on the line energy probed.

\section{Summary and Conclusions}
Analysis of data from dedicated H.E.S.S.~II  observations of 18~h towards the vicinity of the Galactic Centre lead to the 95\%~CL exclusion of the $\langle\sigma v\rangle$ value associated to the 130 GeV excess reported in \cite{bib:4} in the {\it{Fermi}}-LAT data. The likelihood method developed for this study has been successfully applied to estimate for the first time the sensitivity for a DM line search with the five telescope configuration of the H.E.S.S. experiment. New constraints on line-like DM signals have been obtained in the line scan in the energy range between 100~GeV and 2~TeV, bridging the gap between previously reported H.E.S.S. phase I and {\it{Fermi}}-LAT results. The analysis reported here has been performed under the hypothesis of the DM halo centred at the 130 GeV excess position, displaced with respect to the gravitational centre of the Galaxy. Moving the centre of the DM halo to $l = 0, b = 0$ implies a loss of sensitivity by a factor of at least eight for the line search studies. The conclusions about the sensitivity of H.E.S.S. in phase II remain valid for explorations close to the Galactic Centre and the current method will be employed on larger observational datasets in the future.

\section{Acknowledgments}
The support of the Namibian authorities and of the University of Namibia in facilitating the construction and operation of H.E.S.S. is gratefully acknowledged, as is the support by the German Ministry for Education and Research (BMBF), the Max Planck Society, the German Research Foundation (DFG), the French Ministry for Research, the CNRS-IN2P3 and the Astroparticle Interdisciplinary Programme of the CNRS, the U.K. Science and Technology Facilities Council (STFC), the IPNP of the Charles University, the Czech Science Foundation, the Polish Ministry of Science and Higher Education, the South African Department of Science and Technology and National Research Foundation, and by the University of Namibia. We appreciate the excellent work of the technical support staff in Berlin, Durham, Hamburg, Heidelberg, Palaiseau, Paris, Saclay, and in Namibia in the construction and operation of the equipment.

\end{document}